# Self-current induced spin-orbit torque in FeMn/Pt multilayers

Yanjun Xu[1,3], Yumeng Yang[1,2], Kui Yao[2], Baoxi Xu[3], and Yihong Wu[1*]

[1]*Department of Electrical and Computer Engineering, National University of Singapore, 4 Engineering Drive 3, Singapore 117583, Singapore*

[2]*Institute of Materials Research and Engineering, A*STAR (Agency for Science, Technology and Research), 2 Fusionopolis Way, 08-03 Innovis, Singapore 138634, Singapore*

[3]*Data Storage Institute, A*STAR (Agency for Science, Technology and Research), 2 Fusionopolis Way, 08-01 Innovis, Singapore 138634, Singapore*

*Author to whom correspondence should be addressed: elewuyh@nus.edu.sg

Extensive efforts have been devoted to the study of spin-orbit torque in ferromagnetic metal/heavy metal bilayers and exploitation of it for magnetization switching using an in-plane current. As the spin-orbit torque is inversely proportional to the thickness of the ferromagnetic layer, sizable effect has only been realized in bilayers with an ultrathin ferromagnetic layer. Here we demonstrate that, by stacking ultrathin Pt and FeMn alternately, both ferromagnetic properties and current induced spin-orbit torque can be achieved in FeMn/Pt multilayers without any constraint on its total thickness. The critical behavior of these multilayers follows closely three-dimensional Heisenberg model with a finite Curie temperature distribution. The spin torque effective field is about 4 times larger than that of NiFe/Pt bilayer with a same equivalent NiFe thickness. The self-current generated spin torque is able to switch the magnetization reversibly without the need for an external field or a thick heavy metal layer. The removal of both thickness constraint and necessity of using an adjacent heavy metal layer opens new possibilities for exploiting spin-orbit torque for practical applications.

## Introduction

Inverse spin galvanic effect (ISGE) can be exploited to manipulate magnetization of ferromagnetic materials with either bulk or structure inversion asymmetry (SIA)[1-7]. In these material structures, a charge current passing through a ferromagnet (FM)[2,3,8] or FM/heavy metal (HM) heterostructures[4,9-15] generates a non-equilibrium spin density through the ISGE, which in turn exerts a torque on the local magnetization of the FM through either s-d (in the case of a transition metal) or p-d (in the case of dilute magnetic semiconductor) exchange coupling. As the ISGE is originated from spin-orbit coupling (SOC), the resultant torque is referred to as spin-orbit toque (SOT). Unlike spin transfer torque (STT), which requires non-collinear magnetization configurations, the SOT can be realized in structures with a uniform magnetization; this greatly simplifies the structure and device design when investigating and exploiting the SOT effect for spintronics applications.

Although SOC induced spin polarization of electrons has been studied extensively in semiconductors[16-18], the investigations of SOC induced non-equilibrium spin density in ferromagnets and the resultant SOT on local magnetization have only been reported recently. Manchon and Zhang[2] predicted theoretically that, in the presence of a Rashba spin-orbit coupling, the SOT is able to switch the magnetization of magnetic two-dimensional electron gas at a current density of about $10^4$–$10^6$ A/cm$^2$, which is lower than or comparable to the critical current density of typical STT devices. The first experimental observation of SOT was reported by Chernyshov *et al.*[3] for $Ga_{0.94}Mn_{0.06}As$ dilute magnetic semiconductor (DMS) grown epitaxially on GaAs (001) substrate. The compressive strain due to lattice mismatch results in a Dresselhaus-type spin-orbit interaction that is linear in momentum. When a charge current passes through the DMS layer below its Curie temperature, 80K in this case, the resultant SOT was able to switch the magnetization with the assistance of an external field and crystalline anisotropy. The lack of bulk inversion asymmetry (BIA) in transition metal FM has prompted researchers to explore the SOT effect in FM heterostructures with SIA. Miron *et al.*[4] reported the first observation of a current-induced SOT in a thin Co layer sandwiched by a Pt and an AlO$_x$ layer. Due to the asymmetric interfaces with Pt and AlO$_x$, electrons in the Co layer experience a large Rashba effect, leading to sizable current-induced SOT. In addition to the Rashba SOT, spin current from the Pt layer due to spin Hall effect (SHE) also exerts a torque on the FM layer through transferring the spin angular momentum to the local magnetization[19]. To differentiate it from the Rashba SOT, it is also called SHE-SOT. Although the exact mechanism still remains debatable, both types of torques are generally present in the FM/HM bilayers. The former is field-like, while the latter is of anti-damping nature similar to STT. Mathematically, the two types of torques can be modeled by $\vec{T}_{FL} = \tau_{FL}\vec{m}\times(\vec{j}\times\vec{z})$ (field-like) and $\vec{T}_{DL} = \tau_{DL}\vec{m}\times[\vec{m}\times(\vec{j}\times\vec{z})]$ (anti-damping like), respectively, where $\vec{m}$ is the magnetization direction, $\vec{j}$ is the in-plane current density, $\vec{z}$ is the interface normal, and $\tau_{FL}$ and $\tau_{DL}$ are the magnitudes of the field-like and anti-damping like torques, respectively[13,15,20]. To date, the SOT effect has been reported in several FM/HM bilayers with different FMs such as CoFeB[11,13-15,20-22], Co[4,10,23-25], NiFe[12,26] and HMs such as Pt[4,10,12,19], Ta[9,11,13,15,20], and W[21]. An average effective field strength of 4×10$^{-6}$ Oe/(A/cm$^2$) has been obtained, except for the [Pd/Co]$_n$/Ta multilayer[27] which was reported to exhibit a very large effective field strength to current density ratio in the range of 10$^{-5}$ Oe/(A/cm$^2$). In the latter case, the spin Hall current from Ta layer alone is unable to account for the large effective field, indicating possible contributions arising from the Pd/Co interfaces internally, though the exact mechanism is not clear. Despite these efforts, however, so far SOT-induced magnetization switching has only be realized in FM/HM structures with ultrathin FM layers.

Here we report on the observation of both ferromagnetism and SOT effect in [FeMn/Pt]$_n$ multilayers with or without an additional thick Pt layer. This work is inspired by our recent observation of SOT effect in FeMn/Pt bilayers[28] and the earlier report of proximity effect at FeMn/Pt interfaces[29]. By controlling the Pt and FeMn layer thickness, we demonstrate that it is possible to achieve both ferromagnetic properties and SOT effect in [FeMn($t_1$)/Pt($t_2$)]$_n$ multilayers above room temperature (RT), with $t_1$ and $t_2$ in the range of 0.2 nm - 1 nm and 0.4 nm - 0.8 nm, respectively. The field-like effective field ($H_{FL}$) to current density ($j_{mul}$) ratio in standalone [FeMn/Pt]$_n$ multiple layers is about $1\times10^{-6}$ Oe/(A/cm$^2$), which is comparable to those observed in Pt/Co/AlO$_x$ trilayers[13,19]. The addition of a thick Pt layer either at the top or bottom helps increase $H_{FL}/j_{mul}$ to a certain extent but within the same order. We further demonstrate that the magnetization of [FeMn/Pt]$_n$ multilayers can be reversibly switched by the current-induced SOT with or without an additional thick Pt layer. The current density for inducing magnetization switching in a standalone multilayer with a total thickness of 8.2 nm is around $7\times10^5$ A/cm$^2$, which is much lower than that of HM/FM bilayers with similar FM thicknesses[10,19]. The realization of self-current induced magnetization switching in these standalone and thick magnetic layers will open new possibilities for practical applications of SOT-based devices.

## Results

**Magnetic properties.** The magnetic properties of Pt(3)/[FeMn($t_1$)/Pt($t_2$)]$_n$/SiO$_2$/Si (hereafter referred to as Batch A) samples with different layer thicknesses and period were characterized using a Quantum Design VSM by cutting the thin film samples into a size of 2.5 mm × 2 mm. Here the number and symbols inside the parentheses denote the thickness of individual Pt and FeMn layers in *nm*, and *n* is the number of period. Throughout this manuscript the sample sequence starts from the top most layer to the substrate unless otherwise specified. The samples were prepared on $SiO_2$/Si substrates (see details in Methods) and their structural properties were characterized using X-ray diffraction and X-ray photoelectron spectroscopy (see Supplementary Note S1 for details). To facilitate the comparison with electrical measurement results, we fixed $n = 5$ (unless specified otherwise) and varied $t_1$ and $t_2$ systematically to investigate how the magnetic properties depend on the individual layer thickness.

All the multilayers were found to exhibit ferromagnetic properties with in-plane anisotropy (see Supplementary Note S2 for detail discussion). Figure 1a shows a typical example of in-plane and out-of-plane hysteresis loops for the sample with $t_1 = t_2 = 0.6$ nm, measured at 50 K and 300 K, respectively. For this specific sample, the coercivity ($H_c$) decreases from 108 Oe at 50 K to ~ 1 Oe at 300 K, with a saturation magnetization ($M_s$) of 286.8 emu/cm$^3$ at 300 K. Both the small $M_s$ and $H_c$ facilitate SOT-induced magnetization switching with a small current. Figure 1b,c,d shows the saturation magnetization of Pt(3)/[FeMn($t_1$)/Pt($t_2$)]$_n$ multilayers as a function of temperature (the *M-T* curve), with the legend denoting $(t_1, t_2) \times n$ (see Supplementary Fig. S2a,b,c for $t_2$-, $t_1$- and $n$- dependence of $T_C$ and $M_s$, respectively). The *M-T* curves were obtained by first cooling the sample from 300 K to 50 K and then recording the magnetic moment while warming up the sample from 50 K to 380 K with an applied in-plane field of 1000 Oe. The field applied was sufficient to saturate the magnetization in the field direction. Although our VSM only allows us to perform the measurements above 50 K, we have confirmed using a separate system for the (0.6, 0.6)×5 sample that the magnetization below 50 K is almost constant between 10 K and 50 K, as shown in Supplementary Fig. S2d. Figure 1b shows the *M-T* curves of samples with $t_1 = 0.6$ nm, $n = 5$, and $t_2 = 0, 0.1, 0.2, 0.4, 0.6$, and 1 nm, respectively. In the range of $t_2 = 0.1$ nm – 0.6 nm, $T_C$ increases slightly from $t_2 = 0.1$ nm to 0.2 nm, and then drops to about 350 K when $t_2$ increases to 0.6 nm. On the other hand, the $M_s$ at 50 K increases gradually with $t_2$ until saturation at $t_2$ = 0.6 nm from about 587.9 to 795.4 *emu/cm³*. However, $T_C$ drops sharply to about 260 K for both the $t_2$ = 0 and $t_2$ = 1 nm samples. The presence of sizable $M_s$ for the $t_2$ = 0 sample, which is essentially a FeMn(3)/Pt(3) bilayer, below 260 K may be ascribed to two origins: canting of spin sub-lattices of FeMn under an applied field and proximity induced moment in the Pt layer near the FeMn/Pt interface. The former leads to a measurable moment in the applied field direction because of the softening of FeMn spin sublattices at small thickness. The latter is caused by the fact that Pt is at the Stoner threshold to become a ferromagnet which can be polarized through magnetic proximity effect when contacting with a ferromagnet such as Fe, Ni and Co[30-34]. It is reasonable to assume that the same will also happen at the FeMn/Pt interface due to uncompensated spins from FeMn. Our control experiments using FeMn(3)/Au(3) (see Supplementary Fig. S3 for comparison of the *M-H* curves between FeMn(3)/Au(3) and FeMn(3)/Pt(3) bilayer samples) revealed that, although proximity effect is indeed present in FeMn(3)/Pt(3) bilayer, its contribution to magnetic moment is small and the measured moment is dominantly from canting of the FeMn spin sub-lattices. Despite its small contribution to the magnetic moment, the Pt layer inside the multilayer structure plays an important role in enhancing ferromagnetic ordering throughout the multilayers when the Pt thickness is in the range of 0.1 to 0.6 nm. In this thickness range, the proximity effect from both sides of Pt is able to couple with each other and also with the neighboring FeMn layers, leading to global FM ordering in the multilayer. However, when $t_2$ is increased further to 1.0 nm and beyond, the central regions of the individual Pt layers remain un-polarized, hindering ferromagnetic ordering throughout the multilayer. This is the reason why $T_C$ of the $t_2$ = 1 nm sample drops back to the same level of FeMn(3)/Pt(3), but its magnetization is much larger than that of the latter. This is understandable because the multilayer has a larger number of FeMn/Pt interfaces and each of these interfaces will contribute to the net magnetization. We now turn to the $t_1$-dependence of magnetic properties. Figure 1c shows the *M-T* curves of samples with $t_2$ = 0.4 nm, $n$ = 5, and $t_1$ = 0.6, 0.8, and 1 nm, respectively. As can be seen, the $M_s$ at low temperature decreases with increasing $t_1$, but $T_C$ remains almost the same. This suggests that FM ordering weakens when the thickness of FeMn increases. However, unlike the case of increasing $t_2$, the increase of $t_1$ up to 1.0 nm does not lead to a sharp decrease of $T_C$, or in other words, the $T_C$ is mainly determined by the degree of polarization of the Pt layer. The last factor investigated is the total thickness, as shown in Fig. 1d. The decrease of *n* leads to gradual decrease of both $M_s$ and $T_C$. Both the surface and size effect may play a role here since the multilayer is sandwiched between thin Pt layers at both the top and bottom. The former is relevant because when *n* is small, the less polarized top and bottom Pt layer may affect the magnetic properties of the multilayer, leading to reductions of both $M_s$ and $T_C$. On the other hand, the size effect which is common for all magnetic materials, may eventually lead to development of partial paramagnetic phase in these materials, and hence a decrease of both $M_s$ and $T_C$. The $T_C$ of a ferromagnetic thin film can be estimated by scaling analysis, *i.e.*, $T_C(\infty) - T_C(d) \propto d^{-1/\nu}$, where $T_C(\infty)$ and $T_C(d)$ are the Curie temperature of bulk and thin film with a thickness *d*, respectively, and $\nu$ is the critical exponent of the bulk correlation length in the range of 0.5 to 0.705 (Refs [35,36]). The fitting of our data to this equation gives a $\nu$ value of 1.6, which is much larger than values obtained for Ni ($\nu$ = 1) and Gd ($\nu$ = 0.625) thin films[36]. As we will discuss shortly about the *M-T* data, this is presumably caused by the finite distribution of $T_C$ itself in the multilayers.

As the FeMn/Pt multilayers are new, it is of importance to study their critical behavior so as to have a better understanding of their magnetic properties. The *M-T* curve of a ferromagnet generally follows the semi-empirical formula[37] (see Supplementary Note S3 on why this model is preferred over other models):

$$M(T) = M(0)\left[1 - s\left(\frac{T}{T_C}\right)^{3/2} - (1-s)\left(\frac{T}{T_C}\right)^{5/2}\right]^{\beta} \tag{1}$$

where $M(0)$ is the magnetization at zero temperature, $T_C$ is the Curie temperature, $\beta$ is the critical exponent representing the universality class that the material belongs to, and $s$ is a fitting constant. As shown in Supplementary Fig. S4, the $M$-$T$ curves can be fitted reasonably well using this formula except that the $\beta$ values of 0.68-0.9, which are 2-3 times larger than that of bulk ferromagnet (see Supplementary Table S1). The fitting result is very sensitive to $\beta$; in other words, the large $\beta$ value must be a characteristic of the multilayer sample. Although we notice that a large value in the range of 0.7 – 0.89 is typically obtained for surface magnetism[38-41], the multilayers discussed here are different from surface magnetism due to their relatively large thickness. As shown in Fig. 1, $T_C$ of the multilayers depends strongly on the individual thickness; therefore, it is plausible to assume that there is a finite distribution of $T_C$ inside the multilayer due to thickness fluctuation induced by the interface roughness. As shown in Supplementary Fig. S5, the $M$-$T$ curves can be fitted very well, especially in the high-temperature region, by assuming a normal distribution of $T_C$ and using $\beta = 0.365$ for all the samples (note that $\beta = 0.365$ is the critical exponent for $M$-$T$ based on three-dimensional (3D) Heisenberg model). As listed in Supplementary Table S2, the width of $T_C$ distribution agrees very well with the range of $T_C$ observed in Fig. 1 for different samples. These detailed analyses revealed that FeMn/Pt multilayers are 3D ferromagnets with a finite $T_C$ distribution.

**Magnetoresistance and Hall resistance**. Figure 2a,b shows the room temperature magnetoresistance (MR) of four Hall bar devices with the structure Pt(3)/[Pt(0.6)/FeMn(0.6)]$_n$/Ta(3)/SiO$_2$/Si (hereafter referred to as Batch B) with $n$ = 4, 5, and 6, measured by sweeping the field in longitudinal (Fig. 2a) and vertical direction (Fig. 2b), respectively, at a bias current of 1 mA. All these devices have a Pt (3) capping layer and a Ta (3) seed layer. The longitudinal MR of all these devices shows a negative peak at low field with negligible coercivity. The amplitude of the peak remains almost constant while the coercivity increases as temperature decreases (see Supplementary Fig. S6 for temperature dependence of the $n = 5$ sample). Although it is not shown here, the transverse MR shows a positive peak at low field. In contrast to the single peak of longitudinal and transverse MR, the out-of-plane MR shows a characteristic "W" shape below the saturation field (Fig. 2b), which cannot be explained by the conventional anisotropic magnetoresistance (AMR) behavior alone. In order to reveal its origin, we have carried out angular dependence measurement by rotating a constant field of 3000 Oe relative to sample on different planes. The results are shown in Fig. 2c,d for the $n = 6$ sample and the sample with structure of Pt(1)/[FeMn(0.6)/Pt(0.6)]$_6$, respectively. In the figures, $\theta_{xy}$, $\theta_{zy}$ and $\theta_{zx}$ are the angles of field with respect to the $x$, $z$, and $z$ axis, when the field is rotated in the $xy$-, $zy$-, and $zx$-plane, respectively. The results in Fig. 2c,d suggest that both AMR, $\rho = \rho_0 + \Delta\rho_{AMR}(\vec{m} \cdot \vec{j})^2$, and unconventional MR (UCMR), $\rho = \rho_0 - \Delta\rho_{UCMR}[\vec{m} \cdot (\vec{z} \times \vec{j})]^2$, are present in the multilayer samples. Here, $\vec{m}$ and $\vec{j}$ are unit vectors in the directions of the magnetization and the current, respectively, $\vec{z}$ represents the normal vector perpendicular to the plane of the layers, $\rho_0$ is the isotropic longitudinal resistivity, and $\Delta\rho_{AMR}$ ($\Delta\rho_{UCMR}$) represents the size of the AMR (UCMR) effect. Based on these relations, the $\theta_{zy}$ – dependence of MR, if any, is dominated by UCMR as the current (along $x$-axis) is always perpendicular to the magnetization direction during $\theta_{zy}$ sweeping. On the other hand, the $\theta_{zx}$ – dependence of MR is mainly attributed to conventional AMR as $y$-component of magnetization is zero when the field is sufficiently strong to saturate the magnetization in the field-direction. Both AMR and UCMR contribute to the $\theta_{xy}$ – dependence of MR. The small amplitude of MR ($\theta_{zx}$) shown in Fig. 2c,d indicates that the MR shown in Fig. 2a is dominantly originated from UCMR. As shown in Fig. 2c,d, the size of UCMR of the sample with a 1 *nm* Pt capping layer and without any Ta seed layer (0.061%) is comparable to that of the sample with both a 3 *nm* Pt capping and a 3 *nm* Ta seed layer (0.079%). This demonstrates that the observed UCMR is not just from the interfaces with Pt(3) and Ta(3); instead it should mainly come from the multilayer itself. Although both models based on spin-Hall magnetoresistance (SMR)[42] and spin-dependent scattering due to spin-orbit coupling[43] at the FM/HM interface can explain the observed UCMR, we believe that the SMR scenario is more relevant in the multilayer structures. In these samples, the individual Pt layers serves as a source for both SHE and ISHE. The FeMn layer in-between serves as a "spin-current valve", which controls the relative amount of spin currents that can reach a specific Pt layer from the neighboring Pt layers. The reflected and transmitted spin-currents combined entering the specific Pt layer will determine the size of the UCMR. With the presence of both AMR and UCMR, the "W"-shaped MR curves in Fig. 2b can be understood as the competition between the two when there is a slight misalignment of the external field from the vertical direction (See Supplementary Note S4 for more details).

Figure 2e,f shows the dependence of planar Hall resistance (PHR) and anomalous Hall resistance (AHR) on magnetic field in the longitudinal and vertical direction, respectively, for the same set of devices whose MR curves are shown in Fig. 2a,b. PHR and AHR are obtained by dividing the measured planar and anomalous Hall voltage by the current flowing only inside the multilayer instead of the total current. In what follows, a positive current refers to the current following in positive *x*-direction and vice versa. Phenomenologically, the PHR and AHR have a characteristic polar and azimuth angle dependence, *i.e.*, PHR $\propto$ sin$2\varphi$ and AHR $\propto$ cos$\theta$, respectively, where $\varphi$ is the angle between the magnetization and positive current direction and $\theta$ is the angle between the magnetization and the sample normal[12]. The PHE signal shown in Fig. 2e resembles well the PHE curve of a typical FM with a small coercivity. These curves are essentially proportional to the first order derivatives of the MR curves shown in Fig. 2a. On the other hand, the AHE signal increases linearly at low field and saturate at about ±2000 Oe which correlates well with the out-of-plane *M-H* curve in Fig. 1a. The nearly linear increase of the AHE signal from -2000 Oe to 2000 Oe and clear saturation beyond this field range shows that ferromagnetic order is developed throughout the multilayer structure, consistent with the magnetic measurement results.

**Spin-orbit torque**. We now turn to the investigation of current-induced SOT in multilayer devices both with and without an additional 3 nm Pt capping layer. To reduce Joule heating, the current sweeping experiments have been performed using pulsed DC current with a constant duration (5 ms) and duty ratio (2.5%). To ensure good reproducibility, we always started the sweeping from zero current and then gradually increased it to a preset value in both positive and negative directions with a fixed step size. The Hall voltage was recorded using a nano-voltmeter from which PHR was obtained by dividing it with the peak value of pulsed current.

Figure 3a,b,c shows the PHR as a function of current density for devices with structures of (a) Pt(3)/[FeMn(0.6)/Pt(0.6)]$_6$/Ta(3)/SiO$_2$/Si, (b) Pt(3)/[FeMn(0.6)/Pt(0.6)]$_4$/Ta(3)/SiO$_2$/Si, and (c) Pt(1)/[FeMn(0.6)/Pt(0.6)]$_6$/SiO$_2$/Si, respectively. Devices (a) and (b) both have a 3 nm Pt capping layer and a 3 nm Ta seed layer, whereas Device (c) only has a 1 nm Pt capping which is necessary to prevent the sample from oxidation. Due to the large resistivity of Ta as compared to Pt, current passes through the Ta layer can be ignored. To facilitate comparison with Device (c), in Fig. 3a,b, we show the current density in the multilayer in the lower

horizontal axis and the current density in the Pt layer in the upper horizontal axis. The results shown in Fig. 3a,b,c can be reproduced consistently. For the sake of clarity, we only show the result of one round of measurement in which a pulsed current is firstly swept from 0 to a positive preset current (50 mA for (a), 40 mA for (b), and 20 mA for (c)), then to the negative preset current with the same peak value by passing zero, and finally back to zero. The overall shape of the PHR curve can be qualitatively understood if we consider a field-like effective field ($H_{FL}$) induced in the $\vec{z} \times \vec{j}$ direction[12,19,44], as shown schematically in Fig. 3d (top-view of the Hall bar). The current shown in Fig. 3d is the actual current applied to obtain the switching curve in Fig. 3a. Due to the small uniaxial anisotropy, the effective easy axis at the junction of Hall bar is assumed to be at an angle $\alpha$ (e.g., -10°) away from the x-axis. When a current is applied in x-direction, an effective field $H_{FL}$ will be generated in y-direction with its strength proportional to the current. The competition between $H_{FL}$ and the effective anisotropy field ($H_k$) leads to an in-plane rotation of the magnetization to towards y-direction with an angle $\varphi - \alpha$, where $\varphi$ is the angle between the magnetization and x-axis. The PHR reaches the first positive maximum when $\varphi = 45°$. Further increase of the current will rotate the magnetization to a direction that is slightly passing over the y-axis towards the negative x-direction due to the added effect from $H_k$. When the current is gradually reduced after it reaches the positive preset value (50 mA in this case), the magnetization will continue to be rotated in anticlockwise direction and settle down in the opposite direction, i.e., $\varphi = 180° + \alpha$, when the current returns to zero. During this quadrant of sweeping, a negative peak in PHR appears when $\varphi = 135°$. By the same reasoning, the magnetization will continue to be rotated in anticlockwise direction when the current is swept from zero to -50 mA and then back to zero. This is because the effective field direction will be reversed when current changes sign. During this process, the PHR will first reach a positive maximum at $\varphi = 225°$ and then a negative maximum at $\varphi = 315°$. The magnetization will go back to the initial equilibrium direction after a full cycle of current sweeping. Therefore, the results in Fig. 3a,b,c demonstrate clearly that the magnetization of the multilayer device can be switched from one direction to its opposite, and then back to its initial direction (see Supplementary Note S5 for measurements with an additional bias field in y-direction and Supplementary Note S6 for simulation of the PHR curve). It is worth noting that such kind of reversible switching can be realized in a bare multilayer without an additional thick Pt layer, as shown in Fig. 3c. Furthermore, the threshold current density is even smaller than that of the samples with an additional thick Pt layer (Fig. 3a). These results show clearly that an effective field is induced inside the multilayer itself, regardless of whether there is an additional thick HM layer.

Second order PHE measurements[12,45] were then performed to quantify the strength of current-induced effective field $H_{FL}$ in different samples (see details in Supplementary Note S7). Figure 4a shows the $H_{FL}$ for Batch B samples with $n$ = 4, 5 and 6, together with that of the Pt(1)/[FeMn(0.6)/Pt(0.6)]$_6$ and Pt(1)/[FeMn(0.6)/Pt(0.6)]$_4$ samples, which are plotted against the current density in the multilayer portion of the samples ($j_{mul}$). It is worth noting that the effective fields of both Pt(1)/[FeMn(0.6)/Pt(0.6)]$_6$ and Pt(1)/[FeMn(0.6)/Pt(0.6)]$_4$ are comparable with the samples with a thick Pt capping layer, especially at low current density. This shows that the effective field is mostly generated inside the multilayer itself; the effect of spin-current generated by the thick Pt layer is largely confined near its interface with the multilayer. Figure 4b compares the effective field of Pt(1)/[FeMn(0.6)/Pt(0.6)]$_4$/SiO$_2$/Si with that of Pt(3)/NiFe(4.8)/Ta(3)/SiO$_2$/Si trilayer by plotting it against the current density in the multilayer itself for the former and that in Pt layer for the latter. The thickness of the multilayer (excluding the 1 nm Pt capping layer) is intentionally made the same as that of NiFe in the trilayer structure. For the same current density, the effective field of the multilayer is about 4 times larger than that of the trilayer and the difference is even larger if we take into account only the current flowing through the Pt layers. In addition, we have also investigated the $H_{FL}$ for Pt(1)/[FeMn($t_1$)/Pt($t_2$)]$_5$ samples with different FeMn ($t_1$) and Pt ($t_2$) layer thickness combinations. As shown in Supplementary Fig. S11, the $H_{FL}$ increases sharply with the Pt thickness from 0.2 to 0.6 nm with fixed FeMn thickness ($t_1 = 0.6$ nm), which indicates clearly that the spin current is mainly from the Pt layer which itself has already been polarized by the proximity effect. The effect of FeMn thickness is relatively small and is only caused by the difference in uncompensated spins.

**Write and read by current**. To further demonstrate reversible magnetization switching of the multilayer, PHE measurements were performed on Pt(1)/[FeMn(0.6)/Pt(0.6)]$_6$ with alternate write and read pulse as shown schematically in the upper panel of Fig. 5a. The measurement began with the supply of a +20 mA (corresponding to a current density of 1.25 × 10$^6$ A/cm$^2$) write current pulse ($I_w$) with a duration of 5 ms to saturate the magnetization into a specific easy axis direction, followed by reading the Hall voltage with a 5 ms read current pulse ($I_r$) of +2 mA. The reading was repeated 13 times during which the PHR was recorded by dividing each measured Hall voltage with the 2 mA reading current, and the results are shown in the lower panel of Fig. 5a. Subsequent to this, a negative current pulse of -20 mA was applied to reverse the magnetization and then read with the same 2 mA current pulse. The write and read cycles were repeated 8 times, as shown in Fig. 5a. The readout process can be readily understood with the assistance of the schematic diagram in Fig. 5b. During readout, the read current pulse (+2 mA) induces a small rotation of the magnetization ($\delta\varphi$) towards +y direction from its equilibrium positions, one at angle $\alpha$ (State #1) and the other at $\alpha + 180°$ away from +x direction (State #2). When the read current is chosen properly for a specific $a$ value, the magnetization will be rotated to the first octant for State #1 but remains in the second octant for State #2. This leads to Hall resistance of different polarity for the two states, positive for State #1 and negative for State #2. The absolute value of PHR depends on the readout current and misalignment angle $\alpha$, as shown clearly in Fig. 3. The results shown in both Fig. 3 and Fig. 5 demonstrated unambiguously reversible switching of magnetization solely by a current.

## Discussion

Although the physical origin of the field-like effective field in FM/HM hetero-structures is still debatable, recent studies suggest that its ratio to charge current density in the HM layer ($j_c$) can be written in the following form by taking into account only the spin current generated by SHE in the HM layer[46,47]:

$$H_{FL}/j_c = \frac{\hbar}{2e} \frac{\theta_{SH}}{\mu_0 M_s t_{FM}} \frac{g_i}{(1+g_r)^2 + g_i^2} \left(1 - \frac{1}{\cosh(d_{HM}/\lambda_{HM})}\right) \quad (2)$$

where $\theta_{SH}$ is the spin Hall angle of HM, $M_s$ the saturation magnetization of FM, $t_{FM}$ the thickness of FM, $\hbar$ the reduced Planck constant, $e$ the electron charge, $\mu_0$ the vacuum permeability, $d_{HM}$ the thickness of HM, $\lambda_{HM}$ the spin diffusion length in HM, and $g_r = \text{Re}[G_{MIX}]\rho_{HM}\lambda_{HM}\coth(d_{HM}/\lambda_{HM})$, $g_i = \text{Im}[G_{MIX}]\rho_{HM}\lambda_{HM}\coth(d_{HM}/\lambda_{HM})$ with $G_{MIX}$ the spin mixing conductance of FM/HM interface and $\rho_{HM}$ the resistivity of HM. If we use the parameters: $\mu_0 M_s = 0.52$ T for NiFe (much smaller than the bulk value), $\theta_{SH} = 0.2$ (0.004 – 0.34 in literature), $\lambda_{HM} = 1.5$ nm (0.5 nm – 10 nm for Pt in literature), $d_{HM} = 3$ nm, $t_{FM} = 4.8$ nm, $\rho_{Pt} = 31.66$ $\mu\Omega\cdot$cm (measured value), and $G_{MIX} = (8.1 \times 10^{14} + i\, 2.2 \times 10^{14})$ $\Omega^{-1}$m$^{-2}$ for NiFe/Pt[26,48-50], we obtain the field-to-current ratio $H_{FL}/j_c = 1.34 \times 10^{-7}$ Oe/(A/cm$^2$); this is comparable to the experimental value of $2.93 \times 10^{-7}$ Oe/(A/m$^2$) for the Pt(3)/NiFe(4.8)/Ta(3) sample shown in Fig. 4b. However, if we use $d_{HM} = 1$ nm and keep other parameters the same, the effective field to current ratio decreases to $4.0 \times 10^{-8}$ Oe/(A/m$^2$). In other words, if we replace NiFe by the multilayer, the spin current from the 1 nm Pt capping layer alone would be too small to account for the effective field obtained experimentally.

Now the question is: what could be the SOT generation mechanism in the multilayer without an additional thick Pt layer, *e.g.*, in the case of Pt(1)/[FeMn(0.6)/Pt(0.6)]$_6$? The observation of clear SMR suggests that spin current is present inside the multilayer. Considering the fact that FeMn has a very small spin Hall angle[51] and both the Pt and FeMn layers are very thin ($t_2$ is smaller than spin diffusion length of Pt), we may assume that the spin current is dominantly generated in the Pt layers and absorbed almost locally by the uncompensated moment of neighboring FeMn layers (see illustration of spin Hall angle and $M_s$ distribution in Supplementary Fig. S12). Compared with the case of FM/HM bilayers, in which the spin current is generated non-locally due to the large thickness of HM, and the case of bulk material with broken spatial inversion symmetry like strained GaMnAs, in which the spin current is generated locally, the present case falls somewhere between the two; therefore, the SOT observed in FeMn/Pt multilayers can be considered as a pseudo-bulk effect (see Supplementary Note S8 for detailed explanations). In the case of a pure HM layer, when a charge current is applied in *x*-direction, the SHE generates a spin current flowing in *z*-direction with the spin polarization in *y*-direction, thereby building up spin accumulations at both the top and bottom surfaces. At steady state and under the boundary conditions, $j_{sy}^z(0) = j_{sy}^z(d) = 0$, the spin current is given by

$$j_{sy}^z(z) = j_{s0}^{SH} \left[ \sinh\frac{z}{\lambda} - \sinh\frac{z-d}{\lambda} \right] \Big/ \sinh\frac{d}{\lambda} - j_{s0}^{SH} \tag{3}$$

where $j_{s0}^{SH}$ is the SHE spin current, $\lambda$ is the spin diffusion length and *d* thickness of the HM layer. In the case of FeMn/Pt multilayers, in addition to Pt, we also have FeMn layers and the entire multilayer is a ferromagnet. Therefore, the SHE spin current will be partially absorbed and converted to SOT. The absorption is strongest when the polarization of spin current is perpendicular to the magnetization direction and smallest when they are parallel, thereby inducing the SMR-like magnetoresistance. It should be pointed out that in the latter case, spin current can presumably travel through the multilayer because it behaves like a single phase FM, which is different from a FM/HM bilayer. In the extreme case, we may assume that the spin current generated by the Pt layers is completely absorbed by the FeMn layers locally when the polarization of spin current is perpendicular to the local magnetization direction. Under this assumption, there will be no spin accumulation at the two surfaces. The difference in spin current between these two cases gives the SMR-like MR as follows (see Supplementary Note S9 for details):

$$\frac{\Delta R_{xx}}{R_{xx}} = \frac{2\lambda \eta \theta_{SH}^2}{d} \left( \cosh \frac{d}{\lambda} - 1 \right) \bigg/ \sinh \frac{d}{\lambda} \qquad (4)$$

Here, $\eta < 1$ describes the efficiency of spin current absorption in realistic situations. If we use the following parameters: $\eta = 0.5$, $\lambda = 1.5$ nm, $d = 8.2$ nm (total thickness of Pt(1)/[FeMn(0.6)/Pt(0.6)]$_6$), and $\frac{\Delta R}{R_{xx}} = 0.0610\%$ (experimental value extracted from Fig. 2d), we obtain a spin Hall angle $\theta_{SH} = 0.058$ for this sample. With this spin Hall angle, the damping-like effective field to current ratio is calculated as

$$H_{DL}/j_c = \frac{\hbar}{2e} \frac{2\eta \theta_{SH} \lambda}{d \mu_0 M_s t_{FeMn}} \left( \cosh \frac{d}{\lambda} - 1 \right) \bigg/ \sinh \frac{d}{\lambda} \qquad (5)$$

If we use the following parameters: $\mu_0 M_s = 0.32$ T (experimental value), $t_{FeMn} = 3.6$ nm (total thickness of FeMn) and $\theta_{SH} = 0.058$, we have $H_{DL}/j_c = 3.78 \times 10^{-7}$ Oe/(A/cm$^2$). Although it is 2-3 times smaller than the experimentally observed value of $H_{FL}/j_c$, it is a reasonable estimation considering the fact that the field-like and damping-like effective fields are typically on the same order in FM/HM bilayers[13,14,45,52]. It should be pointed out that, although the FeMn layer is sandwiched between two neighboring Pt layers, the top and bottom interfaces are generally different as reported in literature[27,53-55], whereby leading to a net SOT. The degree of asymmetry is represented by the $\eta$ parameter in Eq. (4).

In summary, we have observed both ferromagnetic properties and SOT in FeMn/Pt multilayers consisting of ultrathin Pt and FeMn layers. The former is characterized by a 3D Heisenberg critical behavior with a finite distribution in $T_C$. The self-current induced SOT is able to induce reversible switching of magnetization without the need of an external field and/or additional Pt layer. Such kind of "built-in" SOT in thick films and its ability to switch magnetization without the assistance of an additional HM layer significantly improves the prospects of practical applications of SOT devices.

## Methods

**Sample and experimental geometry**. The FeMn/Pt multilayers consisting of alternate and ultrathin FeMn and Pt layers were deposited on SiO$_2$/Si substrates using DC magnetron sputtering with a base and working pressure of $2\times10^{-8}$ Torr and $3\times10^{-3}$ Torr, respectively. An in-plane field of ~500 Oe was applied during the sputtering deposition to induce a uniaxial anisotropy. The basic structural and magnetic properties of the multilayers were characterized using X-ray diffraction (XRD), X-ray photoelectron spectroscopy (XPS), and vibrating sample magnetometer (VSM), on coupon films. The XRD measurements were performed on D8-Advance Bruker system with Cu K$_\alpha$ radiation. Magnetic measurements were carried out using a Quantum Design vibrating sample magnetometer (VSM) with the samples cut into a size of 2.5 mm × 2 mm. The resolution of the system is better than $6\times10^{-7}$ emu.

The Hall bars, with a central area of 2.3 mm × 0.2 mm and transverse electrodes of 0.1 mm × 0.02 mm, were fabricated using combined techniques of photolithography and sputtering deposition. All electrical measurements (unless specified otherwise) were carried out at room temperature using the Keithley 6221 current source and 2182A nanovoltmeter. The PHE measurements were performed by supplying a DC bias current (*I*) to the Hall bar and measuring the Hall voltage ($V_{xy}$) while sweeping an external field (*H*) in *x*-axis direction. Current sweeping measurements were carried out using pulsed current without any external field.

# References

1.	Ganichev, S. D.*, et al.* Spin-galvanic effect. *Nature* **417**, 153-156 (2002).

2.	Manchon, A., Zhang, S. Theory of nonequilibrium intrinsic spin torque in a single nanomagnet. *Phys. Rev. B* **78**, 212405 (2008).

3.	Chernyshov, A.*, et al.* Evidence for reversible control of magnetization in a ferromagnetic material by means of spin–orbit magnetic field. *Nat. Phys.* **5**, 656-659 (2009).

4.	Miron, I. M.*, et al.* Current-driven spin torque induced by the Rashba effect in a ferromagnetic metal layer. *Nat. Mater.* **9**, 230-234 (2010).

5.	Brataas, A., Kent, A. D., Ohno, H. Current-induced torques in magnetic materials. *Nat. Mater.* **11**, 372-381 (2012).

6.	Locatelli, N., Cros, V., Grollier, J. Spin-torque building blocks. *Nature materials* **13**, 11-20 (2014).

7.	Manchon, A., Koo, H. C., Nitta, J., Frolov, S. M., Duine, R. A. New perspectives for Rashba spin-orbit coupling. *Nat. Mater.* **14**, 871-882 (2015).

8.	Kurebayashi, H.*, et al.* An antidamping spin-orbit torque originating from the Berry curvature. *Nat. Nanotechnol.* **9**, 211-217 (2014).

9.	Suzuki, T.*, et al.* Current-induced effective field in perpendicularly magnetized Ta/CoFeB/MgO wire. *Appl. Phys. Lett.* **98**, 142505 (2011).

10.	Miron, I. M.*, et al.* Perpendicular switching of a single ferromagnetic layer induced by in-plane current injection. *Nature* **476**, 189-193 (2011).

11.	Liu, L.*, et al.* Spin-torque switching with the giant spin Hall effect of tantalum. *Science* **336**, 555-558 (2012).

12.	Fan, X.*, et al.* Observation of the nonlocal spin-orbital effective field. *Nat. commun.* **4**, 1799 (2013).

13. Garello, K., *et al.* Symmetry and magnitude of spin-orbit torques in ferromagnetic heterostructures. *Nat. Nanotechnol.* **8**, 587-593 (2013).

14. Kim, J., *et al.* Layer thickness dependence of the current-induced effective field vector in Ta|CoFeB|MgO. *Nat. Mater.* **12**, 240-245 (2013).

15. Yu, G., *et al.* Switching of perpendicular magnetization by spin-orbit torques in the absence of external magnetic fields. *Nat. Nanotechnol.* **9**, 548-554 (2014).

16. Wunderlich, J., Kaestner, B., Sinova, J., Jungwirth, T. Experimental observation of the spin-Hall effect in a two-dimensional spin-orbit coupled semiconductor system. *Phys. Rev. Lett.* **94**, 047204 (2005).

17. Sih, V., *et al.* Spatial imaging of the spin Hall effect and current-induced polarization in two-dimensional electron gases. *Nat. Phys.* **1**, 31-35 (2005).

18. Valenzuela, S. O., Tinkham, M. Direct electronic measurement of the spin Hall effect. *Nature* **442**, 176-179 (2006).

19. Liu, L., Lee, O. J., Gudmundsen, T. J., Ralph, D. C., Buhrman, R. A. Current-Induced Switching of Perpendicularly Magnetized Magnetic Layers Using Spin Torque from the Spin Hall Effect. *Phys. Rev. Lett.* **109**, 096602 (2012).

20. Avci, C. O., *et al.* Fieldlike and antidamping spin-orbit torques in as-grown and annealed Ta/CoFeB/MgO layers. *Phys. Rev. B* **89**, 214419 (2014).

21. Pai, C.-F., *et al.* Spin transfer torque devices utilizing the giant spin Hall effect of tungsten. *Appl. Phys. Lett.* **101**, 122404 (2012).

22. Qiu, X., *et al.* Spin–orbit-torque engineering via oxygen manipulation. *Nature nanotechnology* **10**, 333-338 (2015).

23. Ryu, K.-S., Thomas, L., Yang, S.-H., Parkin, S. Chiral spin torque at magnetic domain walls. *Nature nanotechnology* **8**, 527-533 (2013).

24. Nistor, C., *et al.* Orbital moment anisotropy of Pt/Co/AlO x heterostructures with strong Rashba interaction. *Physical Review B* **84**, 054464 (2011).

25. Miron, I. M., *et al.* Fast current-induced domain-wall motion controlled by the Rashba effect. *Nature Materials* **10**, 419-423 (2011).

26. Nan, T., *et al.* Comparison of spin-orbit torques and spin pumping across NiFe/Pt and NiFe/Cu/Pt interfaces. *Phys. Rev. B* **91**, 214416 (2015).

27. Jamali, M., *et al.* Spin-orbit torques in Co/Pd multilayer nanowires. *Phys. Rev. Lett.* **111**, 246602 (2013).

28. Yang, Y., *et al.* Fieldlike spin-orbit torque in ultrathin polycrystalline FeMn films. *Phys. Rev. B* **93**, 094402 (2016).

29. Liu, Y., *et al.* Configuration of the uncompensated moments at the FM/AFM interface with a NM spacer. *J. Phys. D: Appl. Phys.* **41**, 205006 (2008).

30. McGuire, T., Aboaf, J., Klokholm, E. Magnetic and transport properties of Co-Pt thin films. *J. Appl. Phys.* **55**, 1951-1953 (1984).

31. Rüegg, S., *et al.* Spin-dependent x-ray absorption in Co/Pt multilayers. *J. Appl. Phys.* **69**, 5655-5657 (1991).

32. Lin, C.-J., *et al.* Magnetic and structural properties of Co/Pt multilayers. *J. Magn. Magn. Mater.* **93**, 194-206 (1991).

33. Poulopoulos, P., *et al.* Magnetic properties of Co-based multilayers with layer-alloyed modulations. *J. Magn. Magn. Mater.* **148**, 78-79 (1995).

34. Emori, S., Beach, G. S. Optimization of out-of-plane magnetized Co/Pt multilayers with resistive buffer layers. *J. Appl. Phys.* **110**, 033919 (2011).

35. Jensen, P., Dreyssé, H., Bennemann, K. Calculation of the film-thickness-dependence of the Curie temperature in thin transition metal films. *Europhys. Lett.* **18**, 463 (1992).

36. Zhang, R., Willis, R. F. Thickness-dependent Curie temperatures of ultrathin magnetic films: effect of the range of spin-spin interactions. *Phys. Rev. Lett.* **86**, 2665 (2001).

37. Kuz'min, M. Shape of temperature dependence of spontaneous magnetization of ferromagnets: quantitative analysis. *Phys. Rev. Lett.* **94**, 107204 (2005).

38. Namikawa, K. LEED Investigation on Temperature Dependence of Sublattice Magnetization of NiO (001) Surface Layers. *J. Phys. Soc. Jpn.* **44**, 165-171 (1978).

39. Alvarado, S., Campagna, M., Hopster, H. Surface magnetism of Ni (100) near the critical region by spin-polarized electron scattering. *Phys. Rev. Lett.* **48**, 51 (1982).

40. Voigt, J.*, et al.* Magnetic hyperfine field at In 111 probes in the topmost atomic layer of Ni (111) surfaces. *Phys. Rev. Lett.* **64**, 2202 (1990).

41. Krech, M. Surface scaling behavior of isotropic Heisenberg systems: Critical exponents, structure factor, and profiles. *Phys. Rev. B* **62**, 6360 (2000).

42. Nakayama, H.*, et al.* Spin Hall magnetoresistance induced by a nonequilibrium proximity effect. *Phys. Rev. Lett.* **110**, 206601 (2013).

43. Zhang, S. S. L., Vignale, G., Zhang, S. Anisotropic magnetoresistance driven by surface spin-orbit scattering. *Phys. Rev. B* **92**, 024412 (2015).

44. Li, H.*, et al.* Intraband and interband spin-orbit torques in noncentrosymmetric ferromagnets. *Phys. Rev. B* **91**, 134402 (2015).

45. Fan, X.*, et al.* Quantifying interface and bulk contributions to spin-orbit torque in magnetic bilayers. *Nat. commun.* **5**, 3042 (2014).

46. Chen, Y.-T.*, et al.* Theory of spin Hall magnetoresistance. *Physical Review B* **87**, 144411 (2013).

47. Kim, J., *et al.* Anomalous temperature dependence of current-induced torques in CoFeB/MgO heterostructures with Ta-based underlayers. *Phys. Rev. B* **89**, 174424 (2014).

48. Weiler, M., *et al.* Experimental test of the spin mixing interface conductivity concept. *Phys. Rev. Lett.* **111**, 176601 (2013).

49. Althammer, M., *et al.* Quantitative study of the spin Hall magnetoresistance in ferromagnetic insulator/normal metal hybrids. *Phys. Rev. B* **87**, 224401 (2013).

50. Vlietstra, N., *et al.* Exchange magnetic field torques in YIG/Pt bilayers observed by the spin-Hall magnetoresistance. *Appl. Phys. Lett.* **103**, 032401 (2013).

51. Zhang, W., *et al.* Spin Hall Effects in Metallic Antiferromagnets. *Phys. Rev. Lett.* **113**, 196602 (2014).

52. Masashi, K., *et al.* Current-Induced Effective Fields Detected by Magnetotrasport Measurements. *Appl. Phys. Express* **6**, 113002 (2013).

53. Bandiera, S., Sousa, R. C., Rodmacq, B., Dieny, B. Asymmetric Interfacial Perpendicular Magnetic Anisotropy in Pt/Co/Pt Trilayers. *IEEE Magnetics Letters* **2**, 3000504-3000504 (2011).

54. Hin Sim, C., Cheng Huang, J., Tran, M., Eason, K. Asymmetry in effective fields of spin-orbit torques in Pt/Co/Pt stacks. *Appl. Phys. Lett.* **104**, 012408 (2014).

55. Huang, K.-F., Wang, D.-S., Lin, H.-H., Lai, C.-H. Engineering spin-orbit torque in Co/Pt multilayers with perpendicular magnetic anisotropy. *Appl. Phys. Lett.* **107**, 232407 (2015).

## Acknowledgements

The authors wish to thank Jingsheng Chen at Department of Materials Science and Engineering and Baoyu Zong at Temasek Laboratories, National University of Singapore, for their help with the XRD and part of the magnetic measurements. We thank Shufeng Zhang at the University of Arizona for helpful discussions. Y.H.W. would like to acknowledge support by the Singapore National Research Foundation, Prime Minister's Office, under its Competitive Research Programme (Grant No. NRF-CRP10-2012-03) and National University of Singapore under its Academic Research Fund (AcRF) Tier 1 (Grant No. R-263-000-A95-112). K.Y. acknowledges the support of IMRE, A*STAR (Agency for Science, Technology and Research) under project IMRE/10-1C0107. Y.H.W. is a member of the Singapore Spintronics Consortium (SG-SPIN).

## Author Contributions

Y.W. devised and supervised the project. Y.X., Y.Y. and Y.W. designed the experiments. Y.X. and Y.Y. performed the experiments (sample fabrication and measurements). Y.W., Y.X., and Y.Y. developed the model and analyzed the data. K.Y. and B.X. participated in the discussion of results. Y.X., Y.Y. and Y.W. wrote the manuscript.

## Additional information

**Competing financial interests:** The authors declare no competing financial interests.

**Figure 1. Magnetic properties.** (**a**) Hysteresis loops of Pt(3)/[FeMn(0.6)/Pt(0.6)]$_5$, measured at 50 K (dashed line in green) and 300 K (dotted line in red) with an in-plane field and at 300 K (black solid curve) with an out-of-plane field. (**b-d**) Saturation magnetization of Batch A samples as a function of temperature (*M-T* curve). The legend *(t$_1$, t$_2$)×n* denotes a multilayer with a FeMn thickness of $t_1$, Pt thickness of $t_2$, and a period of *n*.

**Figure 2. Magnetoresistance and Hall resistance of Batch B samples.** (**a-b**) Magnetoresistance of samples with *n* = 4, 5, and 6, measured by sweeping the field in longitudinal (**a**) and vertical direction (**b**) at a bias current of 1 mA. (**c-d**) Angular dependence of magnetoresistance for the samples with *n* = 6 (**c**) and the sample Pt(1)/[FeMn(0.6)/Pt(0.6)]$_6$/SiO$_2$/Si (**d**), measured by rotating the sample in *xy, zy* and *zx* planes with a constant longitudinal field of 3000 Oe. (**e-f**) planar Hall resistance and anomalous Hall resistance measured by sweeping the field in longitudinal (**e**) and vertical (**f**) direction at a bias current of 1 mA for the same set of samples whose MR curves are shown in Fig. 2a,b. Note that all but the *n* = 6 curve in Fig. 2a,b,e,f are vertically shifted for clarity. The zero-field resistance for samples with *n* = 4, 5, and 6 are 912.6, 871.3 and 769.5 Ω, respectively.

**Figure 3. Current sweeping PHR curves (a, b, c) and illustration of magnetization reversal process (d).** (**a-c**) PHR dependence on the pulsed current density for Pt(3)/[FeMn(0.6)/Pt(0.6)]$_6$/Ta(3)/SiO$_2$/Si (**a**), Pt(3)/[FeMn(0.6)/Pt(0.6)]$_4$/ Ta(3)/SiO$_2$/Si (**b**) and Pt(1)/[FeMn(0.6)/Pt(0.6)]$_6$/SiO$_2$/Si (**c**) samples. Note that $j_{mul}$ represents the current density in the multilayer only, while $j_{Pt}$ represents the current density in the 3 nm Pt layer. (**d**) Schematic illustration of the magnetization switching process assisted by anisotropy misalignment, where *I* represents the total current used in (**a**).

**Figure 4. $H_{FL}$ extracted by second order PHE method.** (**a**) $H_{FL}$ for Batch B samples with $n$ = 4, 5 and 6, together with the Pt(1)/[FeMn(0.6)/Pt(0.6)]$_6$/SiO$_2$/Si and Pt(1)/[FeMn(0.6)/Pt(0.6)]$_4$/SiO$_2$/Si samples. Here, $j_{mul}$ is the current density in the multilayer portion of the samples. (**b**) $H_{FL}$ for Pt(1)/[FeMn(0.6)/Pt(0.6)]$_4$/SiO$_2$/Si and Pt(3)/NiFe(4.8)/Ta(3)/SiO$_2$/Si trilayer. Note that $j_{Pt}$ is the current density inside the 3 nm Pt layer for Pt(3)/NiFe(4.8)/Ta(3)/SiO$_2$/Si.

**Figure 5. Write and read experiment.** (**a**) Illustration of write current pulses (20 mA with a duration of 5 ms) applied to the Pt(1)/[FeMn(0.6)/Pt(0.6)]$_6$ sample (upper panel) and readout signals in terms of PHR (lower panel). Reading is performed with a 2 mA pulse which is repeated 13 times after each writing process. (**b**) Schematic illustration of magnetization rotation during reading at two states with opposite equilibrium magnetization directions.

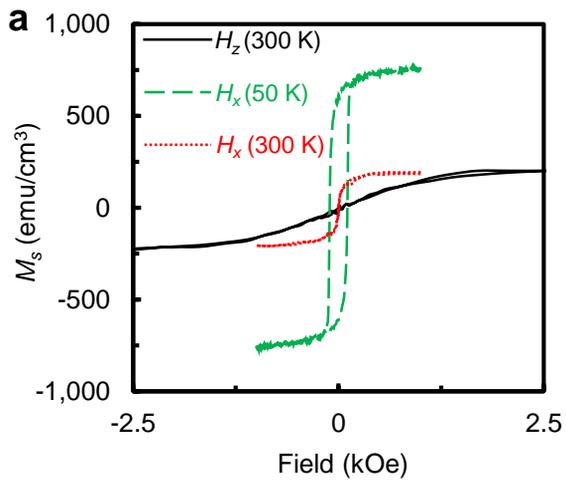
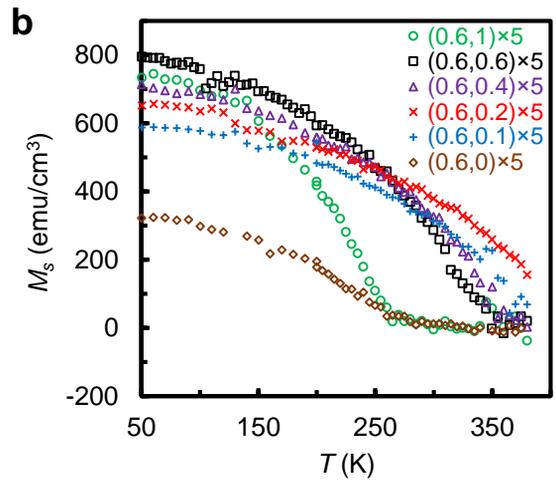
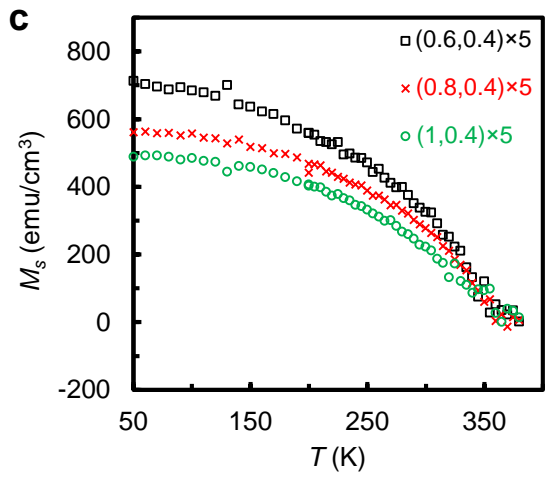
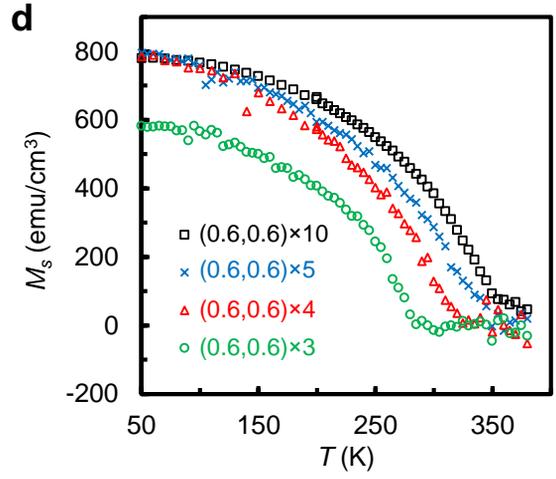

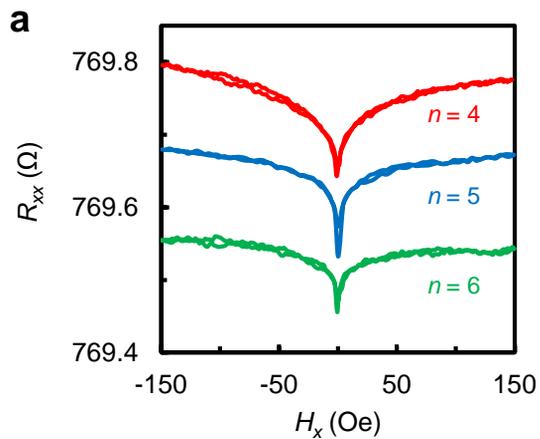
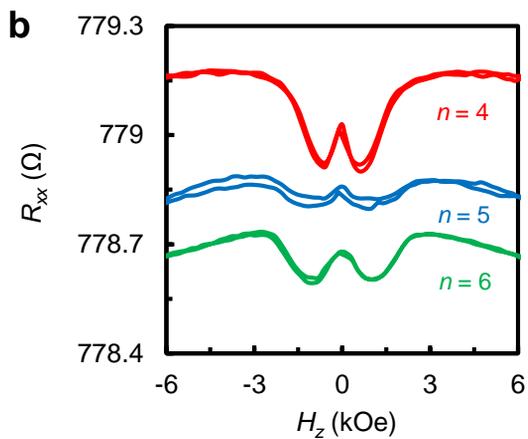
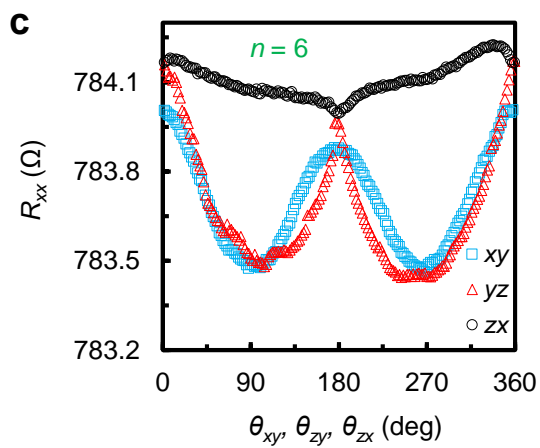
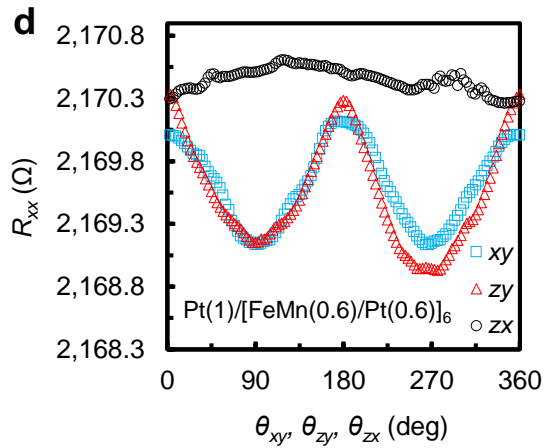
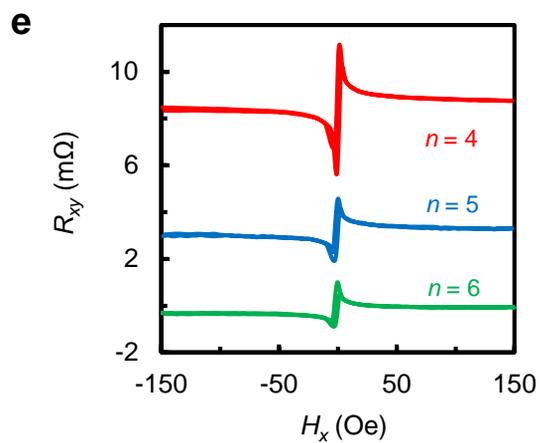
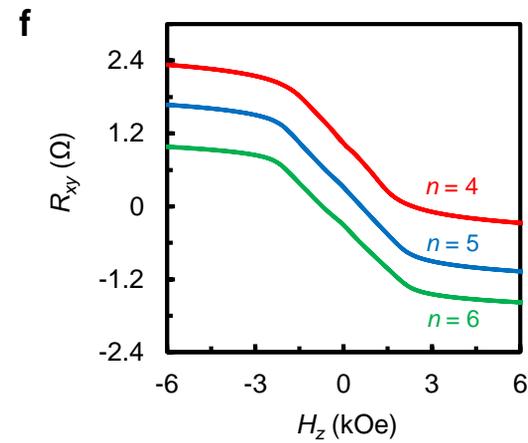

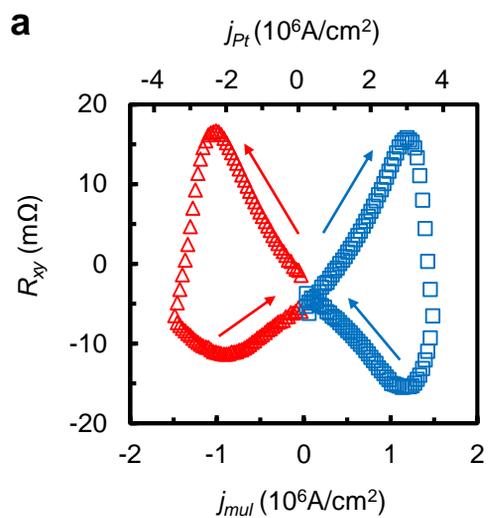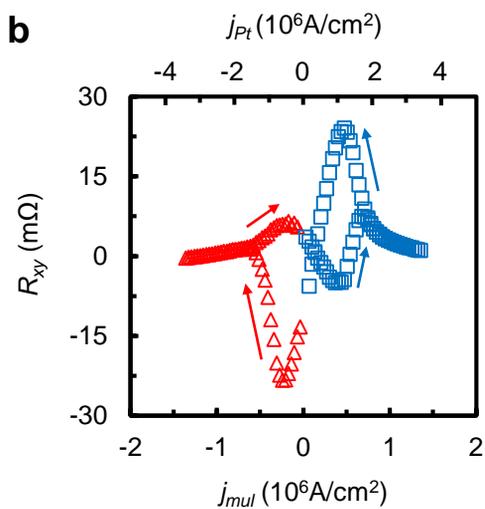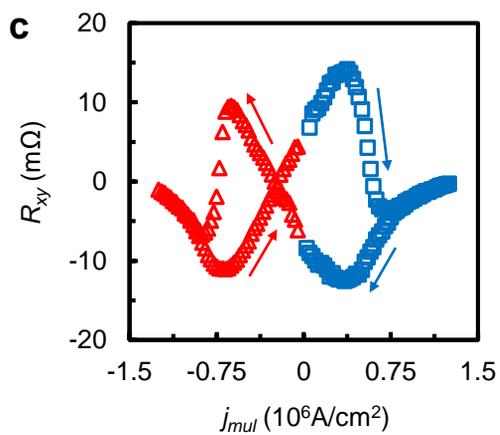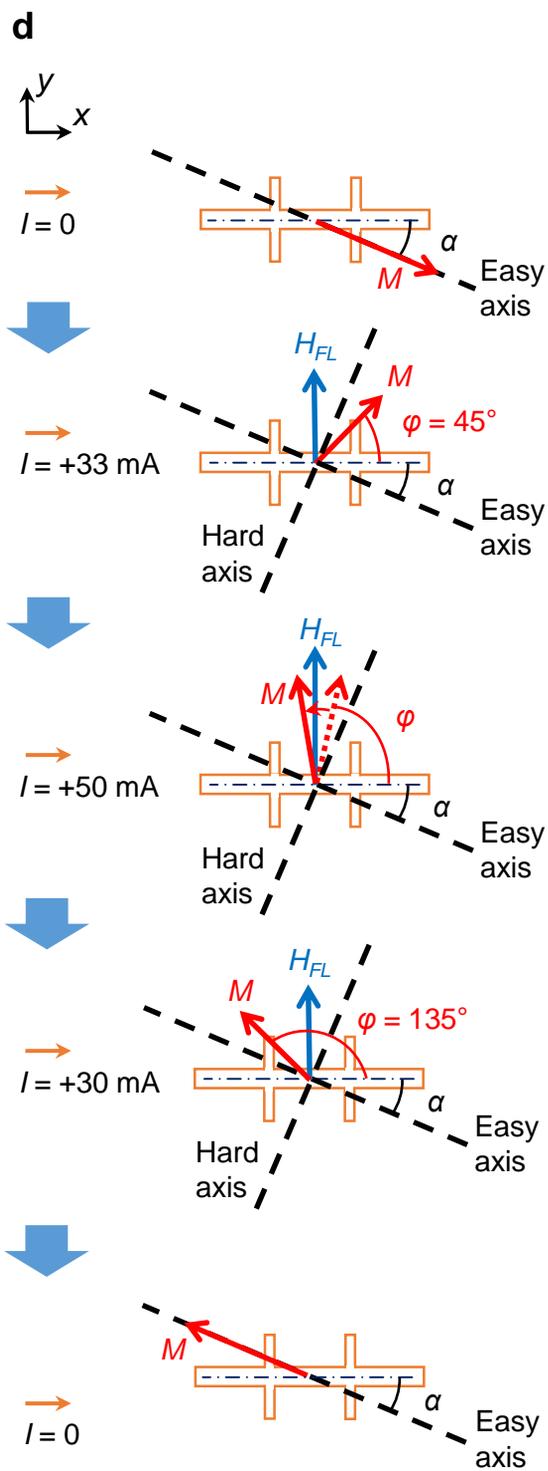

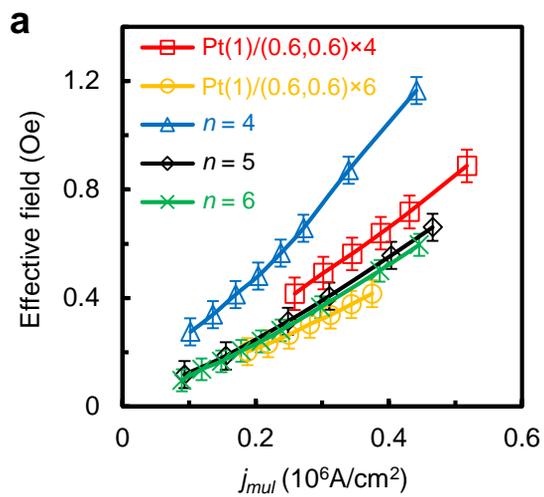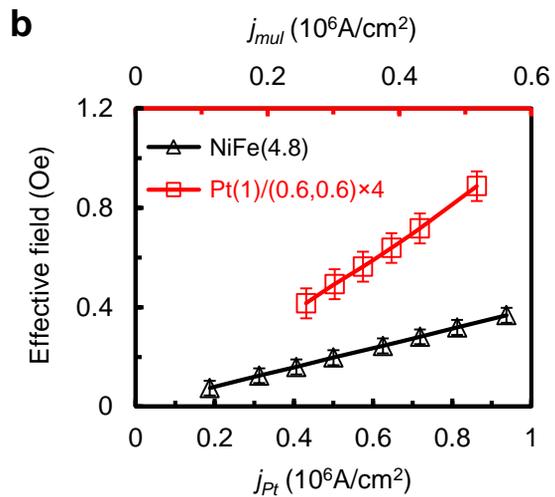

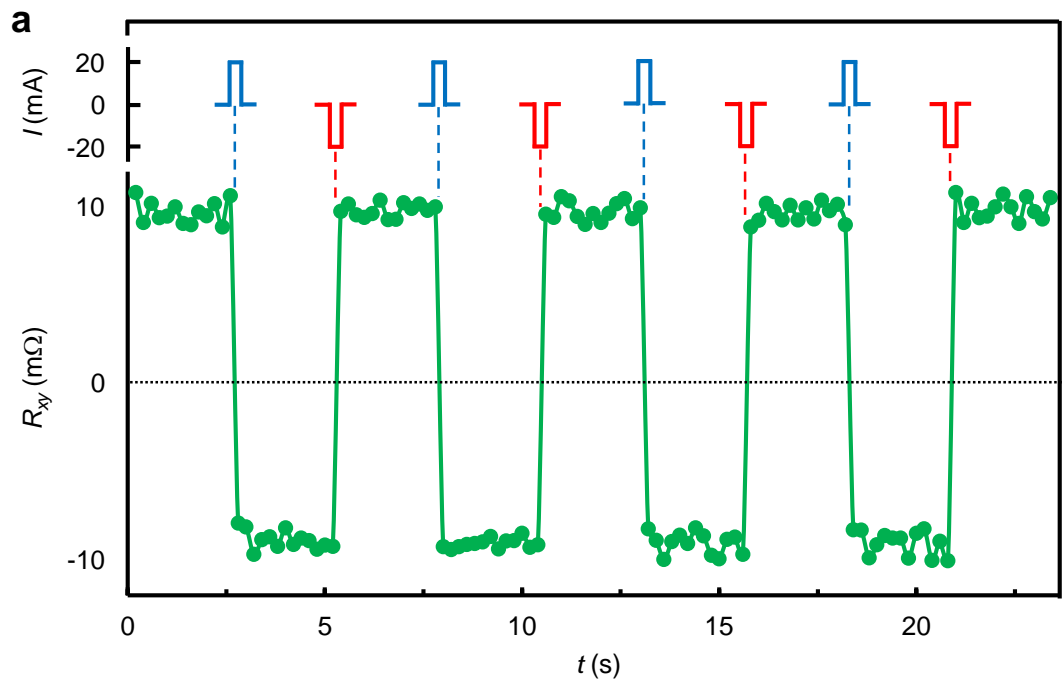

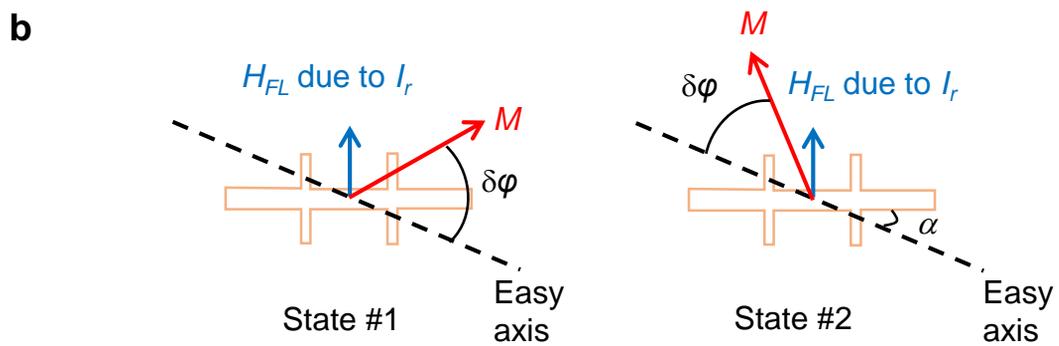